\documentclass{jps-cp}
\usepackage{txfonts} 
\usepackage{graphicx, color}
\usepackage{bm}
\usepackage{physics}
\usepackage{amsmath}
\allowdisplaybreaks[1]

\title{Quantum Transport in Spin-1 Chiral Fermion: Self-Consistent Born Approximation}

\author{Risako \textsc{Kikuchi}$^{1}$, Takumi \textsc{Funato}$^{2,3}$ and Ai \textsc{Yamakage}$^{1}$}

\inst{$^{1}$Department of Physics, Nagoya University, Nagoya 464-8602, Japan \\
$^{2}$Center for Spintronics Research Network, Keio University, Yokohama 223-8522, Japan \\
$^{3}$Kavli Institute for Theoretical Sciences, University of Chinese Academy of Sciences, Beijing, 100190, China}

\email{kikuchi@st.phys.nagoya-u.ac.jp}

\recdate{July 14, 2022}

\abst{Quantum transport for a spin-1 chiral fermion is studied within the self-consistent Born approximation. We find characteristic properties around zero energy, i.e., the peak structure of the density of states and significant suppression of electrical conductivity. These structures originate from the flat band and its interband effect.}

\kword{topological material, spin-1 fermion, quantum transport, self-consistent Born approximation}

\begin{document}
\maketitle

\section{\label{sec:level1}Introduction}

Electronic states in crystals, due to their symmetry, can have high degeneracy, more than twofold, which can emerge a chiral fermion beyond Dirac and Weyl fermions \cite{Beyond2016}. 
The simplest nontrivial case is the threefold degeneracy that yields a spin-1 chiral fermion. 
In recent years, such a state has been diversely discussed in Angle-Resolved Photoemission Spectroscopy (ARPES) measurements of chiral crystal CoSi \cite{Tang2017-kk, Takane2019, Sanchez2019-by, Rao2019-ts, Yuan2019- hz},  
spin transport \cite{Tang21}, optical responses \cite{Flicker18, Sanchez-Martinez2019-ek, Habe2019-ss, Xu2020-zb, Ni2020-ze, Ni2021-eo, Kaushik2021-hx, Dey2022-th}, and spin-1 chiral fermion with quadratic dispersion \cite{Nandy2019-qw, Chen2021-dv, Pal2022-aw}.RhSi, CoGe, and RhGe are in the same cubic lattice with space groups P213 (No.198) as CoSi, have similar electronic states, and are expected to have spin-1 chiral fermions.\cite{Tang2017-kk}

Dirac and Weyl fermions, a prototype of chiral fermion, have been extensively shown to exhibit peculiar transport properties on zero energy \cite{Hosur2013-ck, Armitage2018-sv, Gorbar2020-mi}.
In a two-dimensional massless Dirac fermion, the density of states (DOS) vanishes while the conductivity remains finite on the zero energy, independent of the relaxation time\cite{Fradkin1986-sc, Ludwig1994-ln, Shon1998-ke, Ziegler1998-ar, Tworzydlo2006-hz, Katsnelson2006-yd, Noro2010-cm}. 
A three-dimensional Weyl fermion shows a quantum transport strongly dependent on the impurity concentration \cite{0minato2014, Kobayashi2014, Nandkishore2014-vz, Ominato2015-um, Ominato2016-jl}, giving rise to the semimetal--metal transition.
These exotic quantum phenomena stem from their gapless linear dispersion, called a Dirac cone. 
In contrast, since a spin-1 chiral fermion hosts the flat band in addition to a Dirac cone, it is expected to exhibit unique quantum transport properties.
Remarkable progress has been made in understanding the two-dimensional systems \cite{Vigh2013, Hausler2015-lp, Yang2019-oj, Burgos2022-rl}.

This paper studies the quantum transport of a three-dimensional spin-1 chiral fermion under an impurity potential by the self-consistent Born approximation (SCBA) in the linear response theory. 
We find that the DOS has a peak structure, while the electrical conductivity is significantly suppressed around zero energy.
These low-energy phenomena are attributed to the zero group velocity of the flat band and is mainly due to the interband effect between the flat band and the Dirac cone.

%
%
%

\section{Model}
\label{model}
We consider a three-dimensional spin-1 chiral fermion described by the Hamiltonian
\begin{eqnarray}
\hat{\mathcal{H}}=\hbar v \bm{k} \cdot \hat{\bm{S}},
\end{eqnarray}
where $\bm k$ is the electron wavenumber, $v$ is the Fermi velocity and $\hat{\bm{S}}=(\hat{S}_{x},\hat{S}_{y},\hat{S}_{z})$ are the representation matrices of spin-1 
\begin{align}
\hat{S}_{x} =
\pmqty{
0 & i & 0\\
-i & 0 & 0\\
0 & 0 & 0
},
\
\hat{S}_{y} =
\pmqty{
0 & 0 & -i\\
0 & 0 & 0\\
i & 0 & 0
},
\
\hat{S}_{z} =
\begin{pmatrix}
0 & 0 & 0\\
0 & 0 & i\\
0 & -i & 0\\
\end{pmatrix}.
\end{align}
The eigenenergy is given by $\epsilon_{\lambda,\bm{k}}=\lambda \hbar vk$, where $\lambda$ is the label for the conduction band ($\lambda = 1$), the flat band ($\lambda = 0$), and the valence band ($\lambda = -1$).

As the impurity potential, we assume a Gaussian potential defined by
\begin{align}
\label{gauss}
U(\bm{r}) =  \frac{\pm u_0}{(\sqrt{\pi}d_0)^3}\exp(-\frac{r^2}{d_0^2}),
\end{align}
where $d_0$ is the characteristic length scale and $\pm u_0$ is the strength of the impurity potential. 
The sign $\pm$ means to assume that the numbers of positive and negative valued impurities are the same, so the Fermi level is fixed, irrelevant to the impurity concentration. The Fourier transforms are obtained to be
\begin{align}
 u(\bm{k}) =  \pm u_0\exp(-\frac{k^2}{q_0^2}),
\end{align}
with $q_0 = 2/d_0$.
We also define a dimensionless parameter characterizing the scattering strength
\begin{eqnarray}
W = \frac{q_0n_{\text{i}}u_0^2}{\hbar^2 v^2},
\end{eqnarray}
where $n_{\text{i}}$ is the number of scatterers per unit volume.

\section{\label{result_scba}Density of states}
\begin{figure*}
\includegraphics[width=15cm]{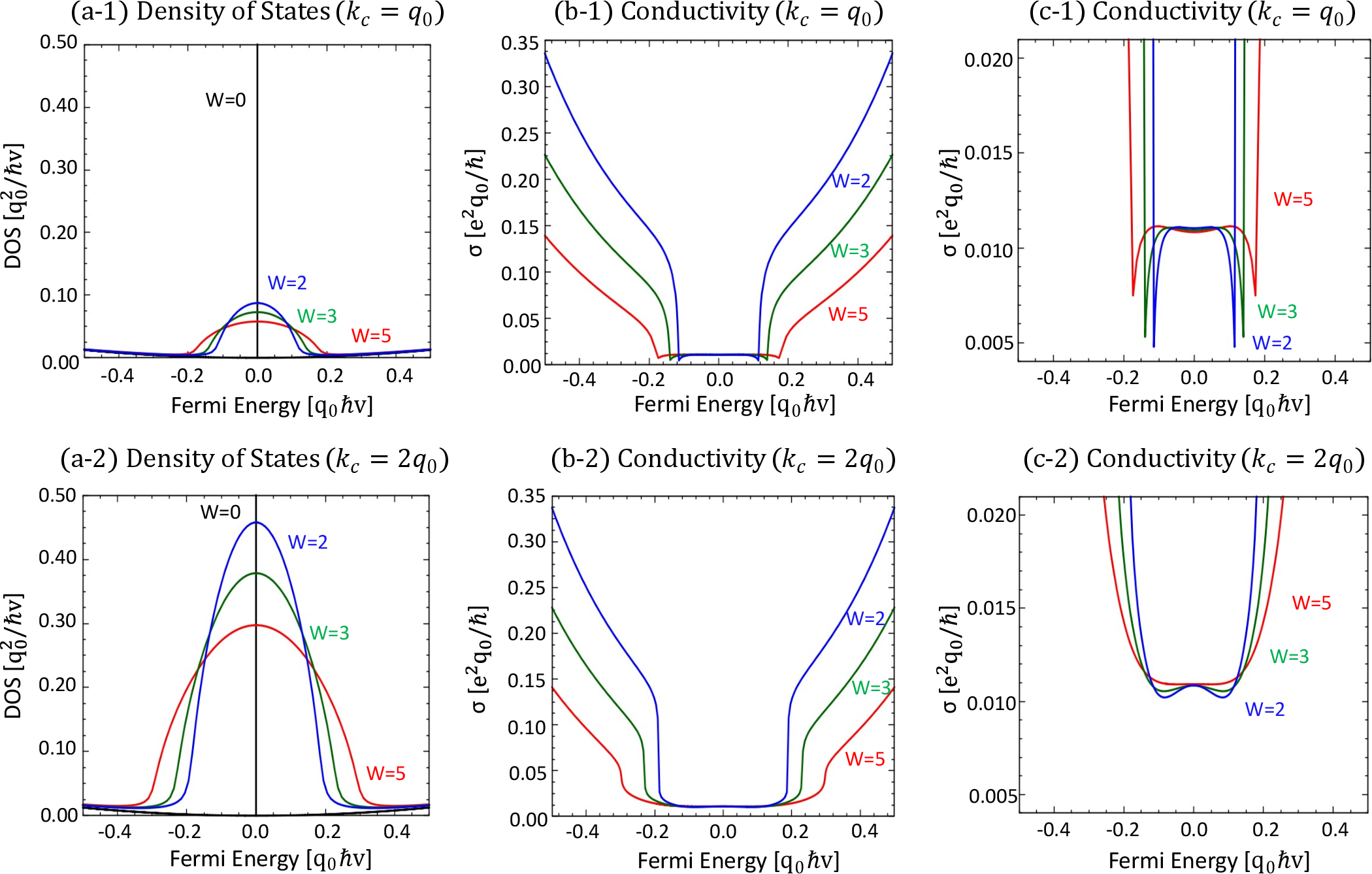}
\caption{Quantum transport for $W=0$ (black line), $W=2$ (blue line), $W=3$ (green line), and $W=5$ (red line). Density of states for (a-1) $k_c=q_0$ and (a-2) $k_c=2q_0$. Conductivity for (b-1) $k_c=q_0$ and (b-2) $k_c=2q_0$. (c-1) and (c-2) are the enlargements near the zero energy of (b-1) and (b-2), respectively.}
\label{impurity}
\end{figure*}
The DOS per unit volume in the clean limit ($W=0$) is given by 
\begin{eqnarray}
D_0(\epsilon)=\frac{\epsilon^2}{2\pi^2(\hbar v)^3}+\frac{k_c^3}{6\pi^2}\delta(\epsilon),
\label{dos0}
\end{eqnarray}
where $k_c$ is the cutoff wavenumber. 
The first term is DOS of the linear dispersion ($\lambda = \pm 1$), and the second term is that of the flat band ($\lambda= 0$) and diverges for $\epsilon = 0$.

Assuming a uniform random distribution of impurities, DOS is obtained from the impurity-averaged Green's function given by 
\begin{align}
\hat{G}(\bm{k},\epsilon +is0) 
= \frac{1}{\epsilon \hat{S}_0
	-\hbar v\bm{k} \cdot \hat{\bm{S}}
	-\hat{\Sigma}(\bm{k},\epsilon+is0)}, 
\label{green function}
\end{align}
where $\hat{S}_0$ is the identity matrix, and the sign $s$ refers to the retarded ($s=1$) and advanced ($s=-1$) Green's functions.
$\hat{\Sigma}(\bm{k},\epsilon)$ is the self-energy matrix approximated in SCBA as
\begin{align}
\hat{\Sigma}(\bm{k},\epsilon+is0)
 = \int\frac{d\bm{k'}}{(2\pi)^3}n_{\text{i}}|u(\bm{k}-\bm{k'})|^2\hat{G}(\bm{k'},\epsilon+is0).
 \label{self energy}
\end{align}
From the obtained Green’s function, DOS per unit volume is calculated as
\begin{align}
D(\epsilon) = -\frac{1}{\pi}\Im\int\frac{d\bm{k}}{(2\pi)^3}\Tr\hat{G}(\bm{k},\epsilon+i0).
\label{dos}
\end{align}

The self-consistent solution is obtained by numerical iteration \cite{Noro2010-cm}.
We discretize the wavenumber as
\begin{align}
	dk_j &=  k_c\frac{j}{\sum_{j=1}^{j_{\text{max}}}j}\label{numerical},
	\quad
	k_j = \frac{1}{2}dk_j + \sum_{j'=1}^{j-1}dk_{j'},
\end{align}
for $j=1,2,...,j_{\text{max}}$ and $k_c$ the cutoff wavenumber. 
Hereafter, we fix $j_{\text{max}}=500$.
The resulting DOS as a function of Fermi energy is shown in Figs.~\ref{impurity}(a-1) and \ref{impurity}(a-2) for $k_c=q_0$ and $k_c=2q_0$, respectively.  
In the high-energy region, the DOS is almost similar to that of the clean limit (the first term of Eq.~(\ref{dos0})) derived from the linear dispersion bands ($\lambda = \pm 1$).
On the other hand, a prominent peak structure can be seen around the zero energy.
This peak originates from a flat band ($\lambda=0$) at $\epsilon=0$ and is broadened by the impurity potential.

As can be seen by comparing Figs.~\ref{impurity}(a-1) and \ref{impurity}(a-2), the results are different for $k_c=q_0$ and $k_c=2q_0$, with the larger $k_c$ indicating a higher and broader peak in DOS. 
Our model assumes a perfectly flat band, which approximates and simplifies the energy bands in the actual material.
Therefore, the cutoff wavenumber $k_c$ corresponds to the range where the band can be approximated as flat. 
As $k_c$ increases, the contribution from the flat band increases, and a larger peak structure appears in DOS.

\section{Conductivity}\label{linear}
The Kubo formula for the conductivity is given by
\begin{align}
\sigma(\epsilon) &= -\frac{\hbar e^2 v^2}{4\pi}\sum _{s,s'=\pm 1}ss'\int \frac{d\bm{k'}}{(2\pi)^3}\text{Tr} [\hat{S}_x\hat{G}(\bm{k'},\epsilon+is0)\hat{J}_x(\bm{k'},\epsilon+is0,\epsilon+is'0)\hat{G}(\bm{k'},\epsilon+is'0)],\label{conductivity}
\end{align}
where $\hat{J}_x(\bm{k},\epsilon,\epsilon ')$ is the current vertex function in the $x$ direction and satisfies the Bethe–Salpeter type equation,
\begin{align}
\hat{J}_x(\bm{k},\epsilon,\epsilon') &=
\hat{S}_x + \int\frac{d\bm{k'}}{(2\pi)^3}n_{\text{i}}|u(\bm{k}-\bm{k'})|^2\hat{G}(\bm{k'},\epsilon)\hat{J}_x(\bm{k'},\epsilon,\epsilon')\hat{G}(\bm{k'},\epsilon').\label{Bethe}
\end{align}

We solve the above equation by the numerical iteration and calculate the conductivity. 
The resulting conductivity as a function of Fermi energy is shown in Figs.~\ref{impurity}(a-1) and \ref{impurity}(a-2) for $k_c=q_0$ and $k_c=2q_0$, respectively. 
In the high-energy region, conductivity increases as $\sigma \propto \epsilon^2$.
Around the zero energy, on the other hand, the conductivity is significantly suppressed.
The energy range of the suppressed area is as broad as the DOS peak.
This suggests that the flat band plays a vital role in conductivity suppression.
Figures~\ref{impurity}(c-1)~and~\ref{impurity}(c-2) show enlarged views of the suppressed area, 
which indicates a conductivity almost independent of $W$ than the non-suppressed area at higher energy.

Let us clarify the details of the characteristic behavior around zero energy and its origin. 
For that purpose, we divide the conductivity into intraband contribution
	\begin{align}
		\sigma_{\text{intra}}(\epsilon) &= -\frac{\hbar e^2 v^2}{4\pi}\sum _{s,s'=\pm1}ss'\int\frac{d\bm{k'}}{(2\pi)^3}
		\bigl(G^s_c G^{s'}_c J^{ss'}_{cc} S_{cc}+G^s_v G^{s'}_v J^{ss'}_{vv} S_{vv} \bigr),
		\label{intra}
	\end{align}
	and interband contribution
	\begin{align}
		\sigma_{\text{inter}}(\epsilon) &= -\frac{\hbar e^2 v^2}{4\pi}\sum _{ss'}ss'\int\frac{d\bm{k'}}{(2\pi)^3}
		\bigl(G^s_c G^{s'}_0 J^{ss'}_{c0} S_{0c}+G^s_0 G^{s'}_c J^{ss'}_{0c} S_{c0}+G^s_v G^{s'}_0 J^{ss'}_{0v} S_{v0}+ G^s_0 G^{s'}_v J^{ss'}_{v0} S_{0v} \bigr),
		\label{inter}
	\end{align}
where the subscripts $c$, $0$, and $v$ denote the conduction, flat, and valence bands in the band basis, respectively.
They are obtained by diagonalizing the Green's function matrix as
	\begin{align}
		\hat U^\dag \hat{\mathcal{H}} \hat U =
		\pmqty{
	  \hbar vk & 0 & 0
	  \\
	  0 & 0 & 0
	  \\
	 0 & 0 & -\hbar vk	
	}, 
\quad
		\hat{U}^{\dagger}\hat{G}(\bm{k},\epsilon+is0)\hat{U}=
		\begin{pmatrix}
			G^s_c & 0 & 0\\
			0 & G^s_0 & 0\\
			0 & 0 & G^s_v\\
		\end{pmatrix}.
	\end{align}
	In this basis, $\hat{S}_x$ is written as
	\begin{eqnarray}
		\hat{U}^{\dagger}\hat{S}_{x}\hat{U}=
		\begin{pmatrix}
			S_{cc} & S_{c0} & 0\\
			S_{0c} & 0 & S_{0v}\\
			0 & S_{v0} & S_{vv}\\
		\end{pmatrix},
	\end{eqnarray}
	and $\hat{J}_x$ is written as
	\begin{eqnarray}
		\hat{U}^{\dagger}\hat{J_{x}}(k, \epsilon+is0, \epsilon+is'0)\hat{U}=
		\begin{pmatrix}
			J^{ss'}_{cc} & J^{ss'}_{c0} & 0\\
			J^{ss'}_{0c} & J^{ss'}_{00} & J^{ss'}_{0v}\\
			0 & J^{ss'}_{v0} & J^{ss'}_{vv}\\
		\end{pmatrix}.
	\end{eqnarray}
	Since the velocity of the flat band is zero, its intraband contribution is zero.
	The interband terms between the conduction and valence bands are also absent. 

\begin{figure*}
\centering
\includegraphics[width=10cm]{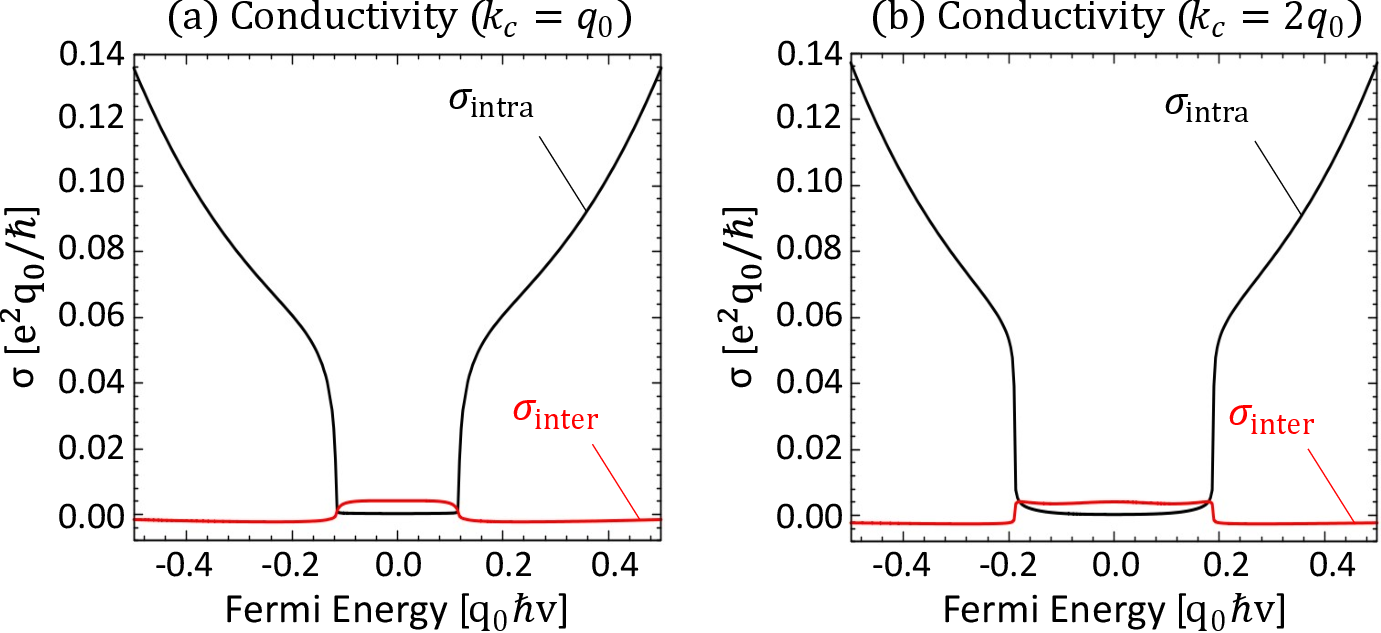}
\caption{
The conductivity from the intraband contribution of the Dirac cone (black line) and from the interband contribution between the Dirac cone and the flat band (red line) for $W=2$, (a) $k_c=q_0$ and (b) $k_c=2q_0$.}
\label{band}
\end{figure*}

Figure~\ref{band} shows the results of the conductivities $\sigma_{\mathrm{intra}}$ and $\sigma_{\mathrm{inter}}$.
The interband term dominates the conductivity around $\epsilon=0$, and the intraband term is nearly zero.

Also, comparing Figs.~\ref{impurity}(b-1)~and~\ref{impurity}(b-2), we find that the larger $k_c$, the wider the energy region where conductivity is suppressed ($\sigma \sim 0.011 e^2 q_0/\hbar$), due to the same reason that the DOS peak broadens with increasing $k_c$.
The conductivity at $\epsilon \sim 0$ is almost independent of $k_c$ and $W$. 
As the density of states increases, the number of scattering channels also increases, and consequently, the interband conductivity is considered to be independent of them.

\section{\label{conclusion}Conclusion}
We studied the quantum transport in a spin-1 chiral fermion system within SCBA, including the current vertex corrections.
We found the peak structure of DOS and the suppression of conductivity around zero energy.
These phenomena originate from the flat band inherent in spin-1 fermions.
We also found that the interband effects between the flat band and the Dirac cone play a vital role in the characteristic behavior, and the conductivity in the vicinity of zero energy is robust against disorder.
The characteristic DOS peak and conductivity suppression cannot be seen in the Born approximation. This is because the Born approximation does not produce a flat band spectral broadening since the self-energy and density of states are proportional to each other. In addition, the vertex correction doubles the intraband effect of the Dirac cone and reduces the interband effect between the Dirac cone and the flat band.
These results provide a basis for elucidating the quantum transport phenomena of spin-1 fermions. Furthermore, nontrivial impurity effects are expected to be latent in chiral fermions beyond Dirac and Weyl fermions, and it will be an exciting problem for quantum transport phenomena in more diverse chiral-fermion systems.

\bibliographystyle{jpsj}
\bibliography{ref}
\clearpage

\end{document}